\documentclass[10pt,letterpaper]{article}
\usepackage{opex3}
\usepackage{lsip_ref}
\usepackage{graphics,graphicx}
\usepackage{caption}
\usepackage{amsmath,amssymb}
\usepackage{cite}

 \newcommand{\rd}[1]{\mathop{\mathrm{d}#1}}
 \newcommand{\der}[2]{\frac{\rd{#1}}{\rd{#2}}}

\begin{document}

%%%%%%%%%%%%%%%%%% title page information %%%%%%%%%%%%%%%%%%
\title{Picosecond all-optical switching in hydrogenated amorphous silicon microring resonators}

\author{Jason S. Pelc,$^\ast$ Kelley~Rivoire, Sonny~Vo, Charles~Santori, David~A.~Fattal, and Raymond~G.~Beausoleil}

\address{Hewlett-Packard Laboratories, 1501 Page Mill Rd., Palo Alto, CA, USA}

\email{$^\ast$ jason.pelc@hp.com} %% email address is required



\begin{abstract}
     We utilize cross-phase modulation to observe all-optical switching in microring resonators fabricated with hydrogenated amorphous silicon (a-Si:H).  Using 2.7-ps pulses from a mode-locked fiber laser in the telecom C-band, we observe optical switching of a cw telecom-band probe with full-width at half-maximum switching times of 14.8~ps, using approximately 720 fJ of energy deposited in the microring.  In comparison with telecom-band optical switching in undoped crystalline silicon microrings, a-Si:H exhibits substantially higher switching speeds due to reduced impact of free-carrier processes.
\end{abstract}


%%%%%%%%%%%%%%%%%%%%%%% References %%%%%%%%%%%%%%%%%%%%%%%%%

%\bibliographystyle{osajnl}
%\bibliography{lsip_nlo,lsip_pic}

%%%%%%%%%%%%%%%%%%%%%%%%%%  body  %%%%%%%%%%%%%%%%%%%%%%%%%%
\section{Introduction}\label{sct:intro}

%Motivation: SiP in computing, nonlinear optics, optical switching devices
As we move to the era of exascale computing systems, the power requirements and performance limitations of electrical interconnects based on metal wires will limit the continued improvement of computing systems as expected by the marketplace and delivered by Moore's law \cite{beausoleil2011lsi}.  Silicon photonics is poised to offer the low power consumption, tight integration with CMOS electronics, and low cost needed to alleviate the strain that interconnection networks are putting on computing systems.  In addition to linear optical components (waveguides, resonators, grating couplers), the optical nonlinearity of silicon provides a means by which some simple logic and switching operations may be carried out in the optical domain without the additional latency and power consumption of optical-electrical-optical conversions \cite{leuthold2010nsp}.

% Intro to a-Si photonics
Recently, hydrogenated amorphous silicon (a-Si:H) has emerged as a candidate for nonlinear optical devices compatible with back-end-of-line (BEOL) CMOS fabrication.  The use of low-temperature deposition processes allows for the possibility of multiple photonic layers integrated with CMOS transistors.  A-Si:H has become more widely used as a photonic material in the past several years as investigators have found that its nonlinear optical properties in the telecom band are superior to those of c-Si.  A-Si:H single- and multimode waveguides were fabricated in 2005 \cite{harke2005lls}, and reports of resonators with $Q$'s exceeding $10^4$ date from 2009, fabricated using both traditional and damascene processes \cite{sun2009tas, selvaraja2009lla}.  Investigations of nonlinear properties have focused either on characterization of material nonlinearity through self-phase modulation experiments \cite{grillet2012asn}, pump-probe experiments to characterize waveguides\cite{kuyken2011npo,narayanan2010bao}, or four-wave mixing for wavelength conversion\cite{matres2013hnf,wang2012upc}.  There have also been recent demonstrations of parametric amplification \cite{kuyken2011ocp} and low-power optical frequency conversion of optical data signals \cite{wang2012upa}.

A central issue inhibiting more widespread use of a-Si:H is the strong dependence of material properties on details of the growth techniques.  Modern a-Si:H growth is done using plasma-enhanced chemical vapor deposition (PECVD).  It has been generally found that material quality improves and defect density decreases with slower growth rates \cite{street2005has}.  As a-Si:H has been studied for many years for its potential in photovoltaic solar cells, a detailed understanding of the role of dangling bond defects has emerged \cite{stutzmann1985lim}.  Dangling bonds result in defect states in the bandgap and could therefore contribute to two-state absorption.  In some nonlinear optics investigations of a-Si:H devices, it has been found that the waveguide properties and nonlinearity degrade over a period of minutes at high optical intensities due to effects that are presumed to be associated with dangling bond defects \cite{kuyken2011npo,kuyken2011ocp}.  Similarly to several recent reports \cite{grillet2012asn,matres2013hnf,wang2012upc}, no optical degradation has been observed in our waveguides over a period of months of investigation, with similar peak pump powers; the a-Si:H growth recipe described in \sct{fab} produces a film which results in stable waveguides.

\subsection{Fundamentals of all-optical switching}

In an optical resonator, if a change $\Delta n$ in the refractive index $n$ can be introduced by optical means, then a frequency shift $\Delta \omega$ in the resonant frequency $\omega_0$ obeys the the relation
\begin{equation}\label{eqn:domega}
  \frac{\Delta \omega}{\omega_0} = -\frac{\Delta n}{n_g},
\end{equation}
where $n_g = n(\omega_0) + \omega_0 \rd{n}/\rd{\omega}|_{\omega_0}$ is the group index.  The imparted resonance shift can bring a probe signal at $\omega_a$ into or out of resonance with the cavity, switching its transmission on or off.  In a microring resonator with a single access waveguide (the all-pass configuration), when a probe signal is on resonance such that $\omega_a = \omega_0$ the probe transmission is low.  When the index change occurs, the probe is then brought out of resonance such that its transmission increases until the index perturbation drops back to zero.

There are several mechanisms by which an optical switch can be driven.  In a material lacking inversion symmetry, the lowest-order bound-electronic optical nonlinearity is due to the $\chi^{(3)}$ tensor.  A common technique, and the one exploited in this paper, is to use cross-phase modulation (XPM), where the index change is proportional to the optical intensity as $\Delta n = n_2 I$.  For Si at 1550 nm, the nonlinear index $n_2 = 4.5\times 10^{-18}$~m$^2$/W \cite{leuthold2010nsp}.  However, due to the Kramers-Kronig relationship arising from very general arguments of causality, nonlinear refraction is always accompanied by nonlinear (two-photon) absorption, which can also give rise to cross-amplitude modulation (XAM) \cite{boyd2008nlo}.  XAM acts to increase the loss $\gamma$ of the optical resonator such that its resonance shift $\Delta \omega$ is also accompanied by an increase in the linewidth $\Delta \gamma$, effectively ``blurring out'' the optical switching effect.  The figure-of-merit for XPM optical switching $\mathrm{FOM} = n_2/(\beta \lambda)$,  where $\beta$ is the two-photon absorption coefficient and $\lambda$ is the wavelength of light, quantifies the quality of an XPM-based optical switch.  The nonlinear FOM has been measured to be between 2 and 5 for a-Si:H \cite{kuyken2011npo, grillet2012asn}, and approximately 0.3 for c-Si \cite{leuthold2010nsp}.

In addition to bound-electronic nonlinear effects due to $\chi^{(3)}$, other physical processes based on real (rather than virtual) excitations can impart an index change $\Delta n$.  Optical switching experiments in c-Si are dominated by effects due to free carriers \cite{almeida2004aoc}.  At a wavelength of 1.5~$\mu$m, two-photon absorption in c-Si or a-Si:H can generate free carriers, which can also cause index changes or absorption changes through the free-carrier dispersion and absorption processes (FCD/FCA), respectively.  As the effects are due to real carriers, the switching speed will be limited by the rate at which carriers recombine or diffuse out of the structure: this speed can be increased by embedding the resonator in a $pin$ diode \cite{turner-foster2010ufc}, or ion implantation to create more recombination sites (at the expense of increased losses) \cite{waldow2008aos}.  Additionally, the absorption of light creates a local temperature shift $\Delta T$, which can act via the thermo-optic effect to create a resonance shift as well.  An important aspect of these two processes is that their speed is determined by the rate at which the physical excitations relax back to their equilibrium values, which can be tens of ns in the case of thermal shifts and between tens of ps to a few ns for carrier effects.

A coupled-mode theory (CMT) model of optical switching in a pump-probe configuration is presented in the Appendix.  A central point is that the combination of the three nonlinear effects described here can be encapsulated in a single nonlinear frequency shift term $\delta \omega^\mathrm{nl}(t)$ included in the CMT equations.  When dynamical equations for the carrier density $N$ and $\Delta T$ are included, a complete picture of the pump--probe interactions is obtained, which we show to agree very well with our observed experimental results.

In \sct{fab} we discuss the techniques for growth of hydrogenated amorphous silicon films and fabrication of optical devices.  In \sct{results} we present results on optical switching experiments performed on microring resonators fabricated out of both a-Si:H and c-Si.  We use a pump--probe technique with pump pulses of duration 2.7~ps, close to the photon lifetime of microring resonators with $Q$'s near $10^4$ at 1.55~$\mu$m of about 8~ps.  We observe switching of the transmission of a probe over a timescale of 14.8~ps, using approximately 720~fJ of deposited energy in a 10-$\mu$m-diameter a-Si:H microring.  The initial experiments on pump--probe switching in undoped c-Si microrings reported switching times of approximately 500~ps \cite{almeida2004aoc}.  Switching times in CMOS-compatible optical switches of 25 ps have been reported in an oxygen-implanted microring \cite{waldow2008aos}, and 135~ps in a polycrystalline Si microring \cite{preston2008hsa}, which were both due to free-carrier effects.

\begin{figure}[t]
  \begin{center}
   \includegraphics[width=\textwidth]{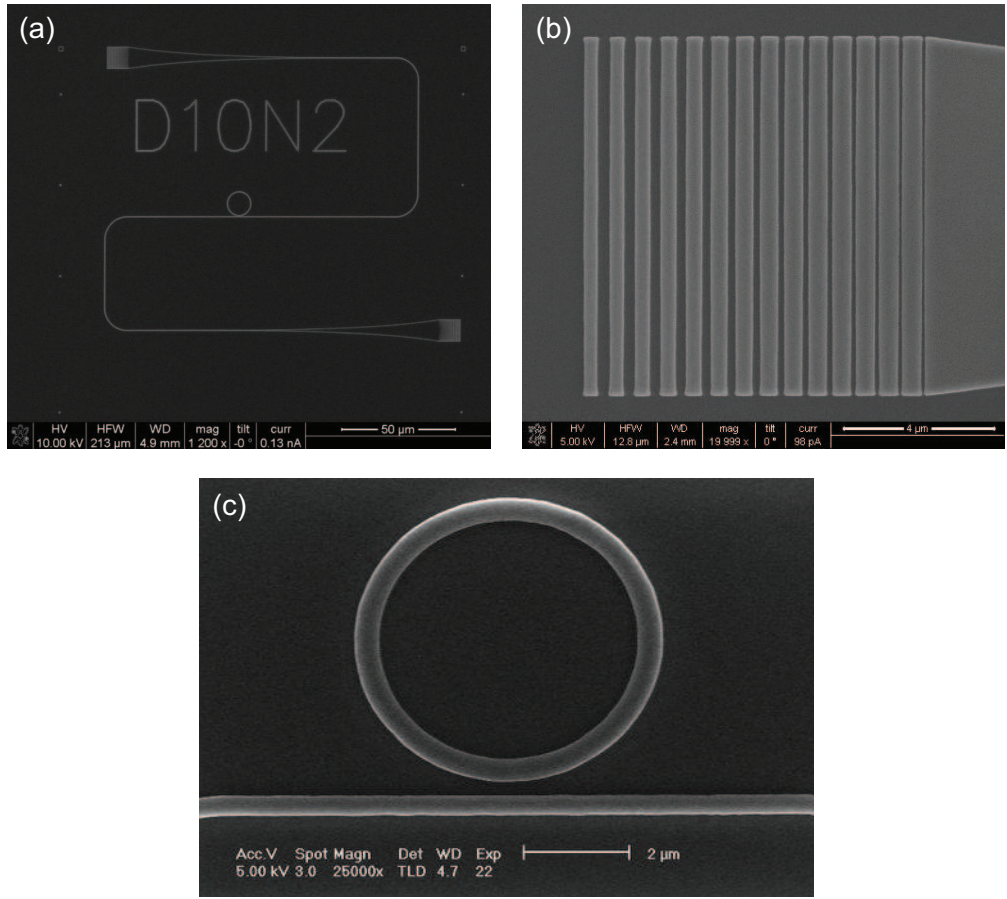}
   \caption{\label{fig:sems} Scanning electron micrographs of a-Si:H devices after etching: (a) layout of microring device with bus waveguide and grating couplers; (b) close-up of grating coupler for TM-polarized light; (c) a 5-$\mu$m-diameter a-Si:H microring and bus waveguide.}
  \end{center}
 \end{figure}

\section{Growth and fabrication}\label{sct:fab}

In this section we give an overview of the a-Si:H growth and device fabrication procedures.  Our fabrication process begins with Si $\langle100\rangle$ wafers upon which 2~$\mu$m of SiO$_2$ has been grown by thermal oxidation.  We then grow an a-Si:H film by plasma-enhanced chemical vapor deposition (PECVD) to the desired thickness.  For the devices described in this paper, the a-Si:H film thickness was 250~nm.  Our PECVD growth process uses SiH$_4$ and Ar gases, flowed at rates of 25 and 50 sccm, respectively, with the chamber maintained at a pressure of 900 mTorr.  The substrate is heated to a temperature of 300$^\circ$~C and two RF tones are used in alternation at a power of 20~W, which we have found improves film quality.  This recipe results in an a-Si:H growth rate of 43 nm/min, and we have been able to grow films of up to 2.5-$\mu$m thickness without delamination.  The refractive index of the film was characterized with ellipsometry to give a result of $n = 3.41$ at 1550~nm, differing from the c-Si value of 3.48 by about 2\%.  We have found that photonic component designs intended for crystalline SOI devices work reasonably well for a-Si:H without any modifications.

Following a-Si:H growth, waveguides are patterned using electron beam lithography using hydrogen silesquioxane (HSQ) resist, followed by reactive ion etching using HBr chemistry.   The a-Si:H film is etched all the way down to the oxide to form wire waveguides with no pedestal layer.  \Fig{sems} shows as-fabricated a-Si:H devices following etching.  After etching, an 800-nm-thick SiO$_2$ cladding layer is grown by PECVD, which provides symmetric optical confinement and passivates the etched surfaces.  In \fig{sems}(a), we see a bent waveguide with a ring in the center of the optical path between two grating couplers.  In \fig{sems}(c), we see a closeup SEM of a 5-$\mu$m-diameter microring and bus waveguide.

The grating couplers use a non-periodic design to minimize Fabry-Perot reflections, and are designed to couple light from a cleaved optical fiber to the TM mode of the a-Si:H wire waveguide.  The grating design is adapted from the design for TE-mode grating couplers presented in \cite{taillaert2004ceb}; an SEM of a fabricated grating is shown in \fig{sems}(b).  TM polarization was chosen because the modes see less sidewall roughness in the tightly bent ring waveguide, which reduces coupling to counter-propagating modes \cite{little1997sri}.  A difficulty with designing grating couplers for TM polarization is that the scattered polarization (in the waveguide) is nearly orthogonal to the input polarization, which reduces the scattering strength of the grating.  Therefore, larger grooves are needed than for TE polarization.  One benefit of the TM grating design is that an additional shallow etch to define the grating grooves is not required; the entire structure including the gratings is fabricated using the same deep etch.  The calculated maximum scattering efficiency for this grating design is approximately -4~dB.

We also fabricated microring resonators and waveguides using a standard undoped crystalline silicon-on-insulator wafer.  The crystalline Si (c-Si) device layer was also 250~nm thick, and was processed the same way.

\section{Setup and results}\label{sct:results}

\begin{figure}[tb]
  \begin{center}
   \includegraphics[width=\textwidth]{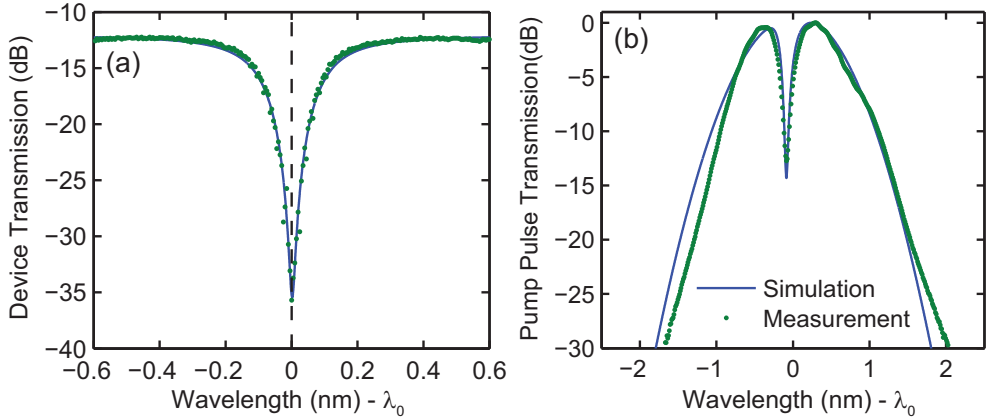}
   \caption{\label{fig:spectra} (a) Measured transmission for 10-$\mu$m-diameter a-Si:H microring using a cw laser scan, and fit to a Lorentzian response.  The observed $Q = 7500$, and extinction ratio is 23.3 dB, showing operation very near the critical coupling point; (b) Spectrum of pump pulses ($P_p = 1.1$~W) transmitted through the ring resonator measured using an OSA, and fit to a CMT simulation with a Gaussian pump pulse with $\tau_p = 2.7$~ps.}
  \end{center}
 \end{figure}

\subsection{Linear spectroscopy of microrings}\label{sct:linear}

We fabricated all-pass (with the geometry shown in \fig{sems}(a)) microring resonators with a range of coupling waveguide widths and coupling gaps and performed laser transmission measurements.  As in c-Si devices, it is found that optimum resonator $Q$ and extinction ratio are observed when the coupling waveguide width is somewhat narrower than the ring waveguide \cite{xu2008smr}.  When measuring the linear optical properties of the microring devices, we use a photonically enabled semiconductor probe station in which we couple light into the devices using a cleaved single-mode fiber, and couple out light using a multimode fiber.  \fig{spectra}(a) shows a measured transmission scan for the 10-$\mu$m a-Si:H microring used in the switching experiments described below.  We have zoomed in on a single resonance at $\lambda_0 = 1554.89$~nm, where we observe $Q = 7500$ and an extinction ratio of $23.3$~dB, showing operation very near critical coupling.  The total throughput of the device was approximately -12~dB, and is attributed mainly to inefficiency of the grating coupler.  When single-mode fibers are used for both excitation and collection from the device, throughput drops to approximately -14.5~dB.  At wavelengths near 1550~nm, the free-spectral range for 10-$\mu$m-diameter (5-$\mu$m-diameter) microrings is found to be approximately 17.1 nm (33.2 nm).  With proper designs (coupling waveguide width between 300 and 400 nm, ring waveguide width 450 nm, and edge-to-edge coupling gap between the ring and bus waveguide between 200 and 350 nm) and using the fabrication process described in \sct{fab}, we obtained $Q$'s of approximately $10^4$ and extinction ratios better than 10~dB, with the best devices having $Q$'s near $2\times 10^4$.

The decay rate $\gamma_0$ of the resonator is important for determining the optical switch speed: $\gamma_0 = \omega_0/Q$.  For the resonator in \fig{spectra}(a), we find $\gamma_0 = 2\pi \times 26$~GHz.  By fitting the microring transmission to theory (solid curve in \fig{spectra}(a)), we can also find the propagation loss of the microring, which we determined to be $\alpha = 8.6$~dB/cm \cite{niehusmann2004uqf}. This loss value is an upper bound for the bus waveguide as we suspect losses in the ring are primarily due to scattering and radiation losses due to the sidewall roughness and the tight bend radius, respectively.  The propagation loss of the bus waveguide does not play a major role in determining the device throughput.  Other researchers have measured propagation losses below 3 dB/cm in a-Si:H wire waveguides using cutback measurements \cite{zhu2012eoc}.

% Nonlinear pump-probe setup
 \begin{figure}[!b]
  \begin{center}
   \includegraphics[width=\textwidth]{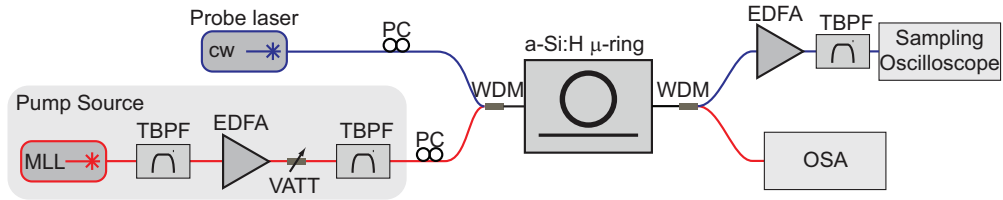}
   \caption{\label{fig:setup} Experimental setup for pump-probe optical switching experiments on microring resonators.  Abbreviations: MLL, mode-locked laser; TBPF, tunable band-pass filter; EDFA, erbium-doped fiber amplifier; VATT, variable optical attenuator; PC, polarization controller; WDM, wavelength-division multiplexer; OSA, optical spectrum analyzer.  The pump path is denoted by red lines, and the probe path denoted by blue; the pump wavelength is positioned one FSR longer than the probe.}
  \end{center}
 \end{figure}

\subsection{Setup for pump--probe switching experiments}\label{sct:setup}

Our experimental setup for demonstration of high-speed optical switching is shown in \fig{setup}.  The pump source is a mode-locked erbium-doped fiber laser which produces pulses of approximately 150 fs duration at a repetition rate of 100 MHz.  We use a tunable band-pass filter to narrow the pulse spectrum to approximately 2 nm and center it on a ring resonance of interest, and then amplify the narrowed pulses with an EDFA, and use a second band-pass filter to clean up the spontaneous emission noise from the EDFA.  Cleaved single-mode fibers are used for both in- and out-coupling from the device for the pump--probe measurements.  The filtered pump source produces pulses of duration $\tau_p = 2.7$~ps ($1/e^2$ intensity radius) with peak powers up to about 50 W in fiber.  The pump is modeled by a Gaussian pulse as $s_\mathrm{in}(t) = s_0 \exp(-t^2/\tau_p^2)$ with the peak power $P_p = |s_0|^2$.  A measured transmitted pump spectrum is shown in \fig{spectra}(b) along with the results of a simulation based on the equations in the Appendix: the pulsewidth $\tau_p$ is extracted from the measured spectrum where we assume a Fourier transform-limited pulse.  The probe source is a cw external-cavity diode laser (ECDL); probe radiation is combined with the pump pulses in a fiber-optic WDM, and is coupled to the devices through the grating coupler.  We position the pump and probe wavelengths on adjacent ring resonances separated by one free-spectral range; this frequency separation allows straightforward combining and separating of the pump and probe signals.  Light transmitted through the device is collected into another single-mode fiber, and sent through a second WDM to separate the pump and probe.  The transmitted pump is sent to an optical spectrum analyzer where we can monitor its spectrum in real time.      The transmitted probe light is amplified with a second EDFA, filtered to remove spontaneous emission noise, and sent to a 55-GHz sampling oscilloscope which is triggered by the pump laser.  In the oscilloscope measurements of probe power versus time, we average 16 traces to reduce noise.  In all experiments, the peak pump power is varied to explore the different physical phenomena, and the cw probe power is chosen to maximize the signal-to-noise ratio without thermally loading the resonator.

\begin{figure}
  \begin{center}
  \includegraphics[width=0.85\textwidth]{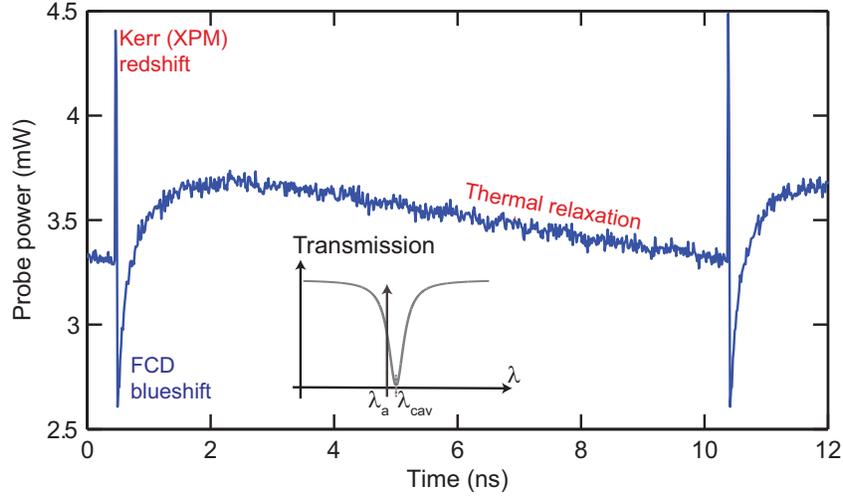}
   \caption{\label{fig:zoomout} Transmitted probe power versus time showing influence of three nonlinear effects: Kerr (XPM) redshift, FCD blueshift, and thermo-optic redshift for an a-Si:H microring resonator with $D = 10$~$\mu$m.  Inset: position of probe wavelength $\lambda_a$ relative to the cavity resonance $\lambda_\mathrm{cav}$ before the arrival of the pump pulse at $t = 0.4$~ns.}
  \end{center}
 \end{figure}

\subsection{Results for a-Si:H microrings}\label{sct:asi-results}

The pulse-to-pulse results, for an a-Si:H 10-$\mu$m-diameter microring, are shown in \fig{zoomout}, where we used a peak pump power $P_p \approx 3$~W.  The inset shows the position of the probe wavelength $\lambda_a$ with respect to the cavity resonance $\lambda_\mathrm{cav}$ at time $t = 0.4$~ns, just before the arrival of a pump pulse.  At $t = 0.4$~ns, the pump pulse arrives, and induces a Kerr-effect (XPM) redshift of the cavity, resulting in a spike in transmission as $\lambda_\mathrm{cav}$ moves further away from $\lambda_a$.  Since the cavity photon lifetime $\tau_0 = 1/\gamma_\mathrm{tot} = Q/\omega_0 \approx 6.2$~ps is longer than the pulsewidth $\tau_p$, it plays the main role in determining the XPM switching timescale by governing the rate at which the pump light exits the resonator.  After the XPM response has died off, the steep drop in transmission is due to FCD caused by carriers created by TPA of the strong pump.  We have measured the free-carrier lifetime in a-Si:H to be approximately 160 ps using these pump--probe experiments (fitting procedures are described in \sct{csi-results}).  Following the diffusion and/or recombination of the carriers, the temperature change due to the heat deposited by pump absorption slowly relaxes until the next pump pulse arrives, 10 ns after the first.  We have also done pump--probe experiments with a pump pulse train with a lower repetition frequency to better characterize the thermal dynamics: we have measured a thermal relaxation time of $\tau_\mathrm{th} = 15.9$ ns for a 10-$\mu$m a-Si:H microring resonator.

\begin{figure}
  \begin{center}
   \includegraphics[width=\textwidth]{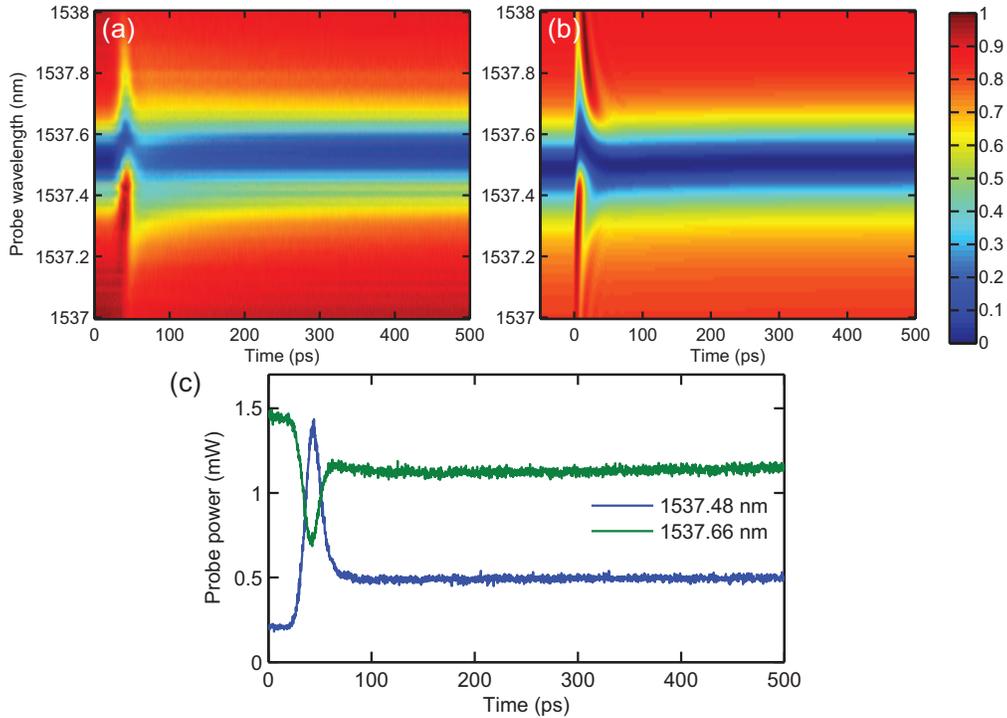}
   \caption{\label{fig:asi} (a) Measured and (b) simulated normalized probe transmission as a function of time and probe wavelength for a-Si:H microring resonator with $D = 10$~$\mu$m.  Red denotes high transmission and blue denotes low transmission. (c) shows individual probe oscilloscope traces when the probe is either on-resonance (blue curve) or red-detuned from the resonance (green curve).}
  \end{center}
\end{figure}

\Fig{asi}(a) shows a series of probe transmission time traces as the probe wavelength was varied.  These probe detuning scans offer an instructive view of the high-speed switching event as one can directly track the temporal evolution of the cavity wavelength as the wavelength at which the measured probe power is lowest.  In this experiment, the peak pump power was set to 1.1~W in the bus waveguide, and the probe power was set to -12.5~dBm (56~$\mu$W).  Because the TPA/FCD effect scales quadratically with the pump power while XPM scales linearly with pump power, in this experiment with approximately $3\times$ less pump power than that for \fig{zoomout}, the carrier effects are reduced by nearly an order of magnitude.  To estimate the energy used in the switching measurement shown in \fig{asi}(a), we note that the peak power of 1.1~W for our fit pulsewidth of 2.7 ps corresponds to a pulse energy of 3.0 pJ.  The spectrum in \fig{spectra}(b) shows that the bandwidth mismatch between the pump and the microring results in incomplete loading of the cavity by the pump pulse.  We define the cavity loading efficiency as the fraction of the incident pump pulse that is dissipated in the cavity.  A cavity loading efficiency of 24\% was observed for the 2.7-ps-pulses by comparing the throughput of the pump pulses whether tuned on or off the cavity resonance, and was confirmed through numerical simulations.  We can use the cavity loading efficiency to determine the energy dissipated in the switching operation, which is found to be 720~fJ.

We used the experimental results of \fig{asi}(a) to help fit the material parameters by comparing with numerical integration of the CMT equations as described in the Appendix.  We assume the grating couplers perform symmetrically and the throughput of the device is dominated by the grating couplers (i.e. propagation loss is zero) to determine $P_p$ in the bus waveguides.  For properties of the microring resonator, we use values determined from linear optics measurements as described in \sct{linear}.  The complexity of the detuning scan makes regression of material parameters very difficult.  By adjusting parameters by hand and simulating a set of probe transmission traces we produce the detuning scan shown in \fig{asi}(b); our parameter values are shown in \tab{params}.

Two probe transmission traces are shown in \fig{asi}(c) for probe wavelengths either very near (1537.48 nm) or slightly red-detuned (1537.66 nm) from the cavity resonance.  When the probe is initially on resonance, its transmission is low until the pump pulse arrives and redshifts the cavity resonance.  Once the pump leaves the cavity, the cavity recovers nearly to its original position, with a slight redshift remaining due to heat deposited in the cavity by absorption of the pump.  In the red-detuned case, the initially non-resonant probe has high transmission until the pump XPM-induced redshift brings the cavity into resonance, causing a probe transmission drop.  The shortest FWHM switching time we observed in the set of traces in \fig{asi}(a) is 14.8 ps.  In both experiments and simulations the switching time is found to decrease with higher incident pulse energy, at the expense of higher carrier and thermal effects.

\begin{figure}
  \begin{center}
   \includegraphics[width=\textwidth]{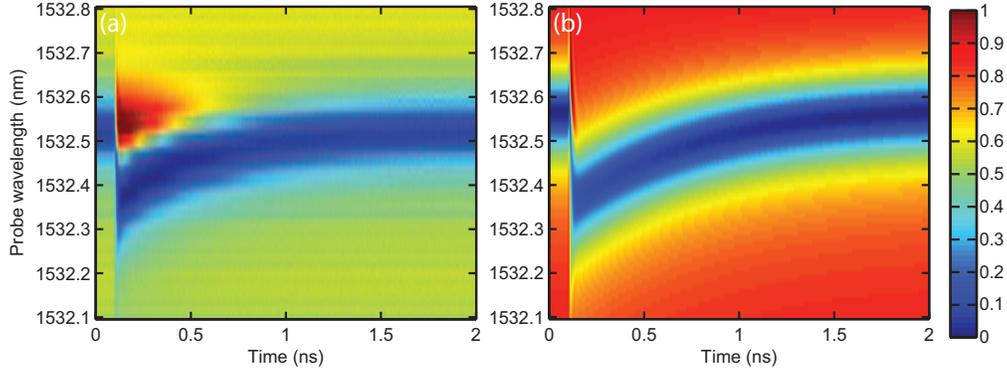}
   \caption{\label{fig:csi} (a) Measured and (b) simulated probe transmission as a function of time and probe wavelength for c-Si microring resonator with $D = 5$~$\mu$m.}
  \end{center}
\end{figure}

\subsection{Results for c-Si microrings}\label{sct:csi-results}

As described in \sct{fab}, we also fabricated c-Si microrings to compare their optical switching performance with those made with a-Si:H.  Our experimental results, for a 5-$\mu$m-diameter ring, are shown in \fig{csi}(a).  In this experiment, we used a pulse energy of 11 pJ for the microring which was observed to have $Q = 8500$ with an extinction ratio of 16~dB.  The pump was positioned on an adjacent resonance at 1565.75~nm.

The experimental results in \fig{csi}(a) indicate that the primary nonlinear effect is a free-carrier-induced blueshift of the cavity.  Although XPM is visible at the time of the pump pulse, its effects are quickly dominated by FCD due to carriers created by TPA.  \Fig{csi}(b) shows simulation results in which we have fit the free-carrier lifetime $\tau_c$ and found it to be 700 ps.  Fitting free-carrier lifetimes for both c-Si and a-Si:H was done using the following technique: from detuning scan data as in \fig{csi}(a) we extract the cavity resonance at each point in time by finding the wavelength with minimum transmission.  We then fit the relaxation of the cavity resonance following the pump pulse to a falling exponential from which we extract the carrier lifetime $\tau_\mathrm{car}$.  A similar procedure was used for measurement of $\tau_\mathrm{th}$.  One difference we observe between the experiment and simulation is that for probe wavelengths initially on the cavity resonance, the switching pulse is higher than the quiescent non-resonant probe power (red area of \fig{csi}(a)).  This is due to a nonlinearity in the EDFA amplification when its input power is very low.

Investigating the switching dynamics in this experiment for an on-resonance probe, we find that the FWHM switching time is 308~ps, a factor of 21 longer than for the a-Si:H results shown in \fig{asi}.  It is possible to reduce the switching time by either embedding the waveguide in a reverse-biased $pin$ diode or implanting oxygen ions \cite{turner-foster2010ufc,waldow2008aos}.

\section{Discussion}\label{sct:disc}

We have fabricated microring resonators using an a-Si:H film deposited at 300$^\circ$~C and used them for demonstrations of high-speed optical switching in the 1.5-$\mu$m telecom band.  In recent years, several research groups have demonstrated superior nonlinear optical properties of a-Si:H with respect to c-Si in the 1.5-$\mu$m telecom band, and have used this to demonstrate ultralow power wavelength conversion \cite{wang2012upa}, and spectral broading due to self-phase modulation \cite{grillet2012asn}.  All-optical switching has been demonstrated in c-Si microring resonators \cite{almeida2004aoc}, but, to the best of our knowledge, it has not been reported in integrated a-Si:H microresonators of any kind.  We note a recent report of XPM-based optical switching in a microcylindrical a-Si:H resonator formed by wet etching of an optical fiber with an a-Si:H core \cite{vukovic2013uoc}. Optical switching in deposited polycrystalline Si microring resonators was reported in 2008 based on free-carrier effects \cite{preston2008hsa}; a-Si:H begins to crystallize when processed at temperatures higher than 350~$^\circ$C, leading to considerably higher waveguide losses \cite{zhu2012eoc}, and nonlinear properties more like bulk c-Si than a-Si:H.  Other CMOS-compatible nonlinear optics platforms include Hydex and Si$_3$N$_4$, but these have substantially lower index contrast and nonlinearity, and are generally not appropriate for optical switching but have interesting applications in wavelength conversion and other parametric processes \cite{Moss2013}.

In our experiments we observe optical switching behavior in a-Si:H microrings to be dominated by cross-phase modulation due to the Kerr effect, while in c-Si devices it is dominated by free-carrier dispersion due to carriers created by two-photon absorption.  This is understood by considering the magnitudes of the relevant coefficients for a-Si:H and c-Si as presented in \tab{params} in the Appendix: for a-Si:H (c-Si) the nonlinear index $n_2$ is $1.0\times10^{-17}$ ($4.5\times 10^{-18}$) m$^2$/W while the effects of FCD are determined by the product of the FCD coefficient $dn/dN$ and the two-photon aborption coefficient $\beta$, which takes the value $dn/dN \times \beta = -2.4\times 10^{-39}$ $(-1.5\times 10^{-38})$~m$^4$/W for a-Si:H (c-Si).  If we assume that $dn/dN$ is equal for c-Si and a-Si:H, this implies that $\beta$ is $1.4\times10^{-12}$ m/W for a-Si:H, from which we calculate a corresponding figure-of-merit for nonlinear switching of 4.7, approximately a factor of 15 higher than for c-Si, and in line with recent experimental demonstrations \cite{grillet2012asn,matres2013hnf}.

In future on-chip photonic networks, it is unlikely that data rates will be required to exceed double the clock frequency, which is expected to stay below 5~GHz into the next decade \cite{ref:georgas2011ald,beausoleil2013paf}.  As such, while we believe the speed demonstrated in this paper which could support data rates as high as 50 GHz, a more compelling research goal would be a significant reduction in switching energy.  Here we have demonstrated a switching energy of 720 fJ.  A calculation of the energy $U$ required to shift the resonance of an optical resonator with normalized mode volume $\tilde{V} = V/(\lambda/n)^3$, linear index $n = n_g$, and nonlinear index $n_2$ by one linewidth (corresponding to a switching extinction ratio of 7~dB) is, using \eqn{domega} \cite{barclay2005nro}:
\begin{equation}\label{eqn:U}
U = \frac{\lambda^3}{n n_2 c} \frac{\tilde{V}}{Q}.
\end{equation}
To achieve lower switching energy, it is therefore desirable to find a material with as high an $n_2$ as possible (with high FOM) while similarly maximizing $Q/\tilde{V}$.  While microrings with smaller mode volume have been demonstrated \cite{xu2008smr}, photonic crystal nanocavities are a superior choice to minimize switching energy due to the fact that their  $Q/\tilde{V}$ can reach nearly $10^6$ ($Q \approx 1.2\times10^6$ and $\tilde{V} \approx 1.6$) \cite{tanabe2007lpd}.  In our $D = 5$~$\mu$m microring resonators, with $Q \approx 10^4$ and $\tilde{V} \approx 10$, it is potentially possible to reduce the switching energy to below 1 fJ, a value on target with state-of-the-art optical switches in III-V nanocavities \cite{nozaki2010sfa}.  An interesting potential benefit of Kerr-based optical switching is that there is a very clear speed/energy tradeoff which can be exploited in a system design: switching energy decreases with increasing $Q$, but the cavity photon lifetime (and therefore switching speed) is proportional to $Q$.

There is uncertainty over why the nonlinear properties of a-Si:H should be so much better than c-Si.  Under the assumption that both materials are indirect gap, if we compare the bandgap energies $E_g$ and apply the theory of Ref.~\cite{sheik-bahae1991dob} for direct gap semiconductors in which the value of $n_2$ is proportional to $E_g^{-4}$, we would expect $n_2$ for a-Si:H to be approximately 20\% of the c-Si value.  However, in the same model it is also found that $n_2$ vanishes when $x = \hbar \omega/E_g \approx 0.74$ and is maximized for $x \approx 1/2$.  At 1550 nm, in c-Si $x = 0.72$ and a-Si:H $x = 0.47$.  When these factors are combined, the $n_2$ ratio for a-Si:H to c-Si is estimated to be approximately 2.5; our measurements of a-Si:H indicate a ratio of approximately 2.3.  However, we note that in measurements of $n_2$ and $\beta$ for c-Si, a sign change of $n_2$ has not been observed, possibly due to contributions to $n_2$ from Raman or quadratic Stark effects \cite{bristow2007tpa}.  It has also been suggested to us that the Si--H polar bonds in a-Si:H may play a role in determining the nonlinear optical properties of the material \cite{vanstryland}.  It remains for future work to investigate the role of hydrogenation on nonlinear optical properties.

\section*{Appendix: Coupled-mode theory model and material parameters}\label{app:cmt}

We model nonlinear dynamics in the ring using a coupled-mode theory (CMT) approach, in which electric fields in the ring are averaged over the volume of the ring.  As such, the field, at a single wavelength, inside the ring resonator is given by a single complex-valued function of time.  For simulations involving the interaction of light at two wavelengths (the pump and probe), we shall simply use two separate variables.  For a single frequency $\omega_a$, our dynamical variable is the field amplitude $a(t)$, which is normalized such that $|a(t)|^2$ is the energy stored in the ring resonator at time $t$.  The equation of motion for $a(t)$ in a ring with no nonlinearity is given by
\begin{equation}\label{eqn:eom-no-nl}
\der{a}{t} = \left[ i \left( \omega_0 - \omega_a\right) -\frac{\gamma_\mathrm{0}}{2} \right] a + \kappa s_\mathrm{a}(t),
\end{equation}
where $\omega_0$ and $\gamma_0$ are the resonant frequency and total decay rates of the cavity, and $s_a(t)$ is the input signal to the cavity, normalized such that $|s_a(t)|^2$ is the optical power in the waveguide \cite{little1997mrc}.  The decay rate is the sum of contributions from coupling to the external waveguide $\gamma_c$, and internal losses $\gamma_l$ and $\gamma_a$ due to scattering (radiation), and absorption, respectively: $\gamma_0 = \gamma_l + \gamma_a + \gamma_c$.
We define the absorption loss fraction as the ratio $\eta_\mathrm{lin} = \gamma_a/(\gamma_a + \gamma_l)$ of the absorption loss to the total internal loss, which is relevant for thermal effects.  By conservation-of-energy arguments, the coupling coefficient between the waveguide field and the cavity field can be found as $\kappa = i(\gamma_c e^{i\phi})^{1/2}$, where $\phi$ is the phase accumulated in the coupling section.  For a cw input $s_a(t) = s_{a0}$, the steady state response of \eqn{eom-no-nl} is the Lorentzian spectral response of an optical cavity.  Critical coupling is defined as the condition where the external loss rate due to the coupling equals the internal loss rate: $\gamma_c = \gamma_l + \gamma_a$.

\begin{table}[ht]
\caption{Parameters describing optical and material properties in amorphous and crystalline silicon microring resonators, and their values.  Values marked with an asterisk were measured in this work, and those not measured in this work are quoted from \cite{van_vaerenbergh2012cem}, except where indicated.  $D$ is the diameter of the resonator.}
\centering
\begin{tabular}{c|l|c|c|l}
\hline\hline
Variable & Description                          & c-Si value & a-Si:H value & units \\
\hline
$E_g$ & Bandgap energy & 1.1 \cite{leuthold2010nsp} & 1.7 \cite{street2005has} & eV \\
$n_2$   & Kerr nonlinear index                  & $4.5\times 10^{-18}$ \cite{leuthold2010nsp}  & $1.0 \times 10^{-17} \ast$ & m$^2$/W \\
$\beta$ & TPA coefficient     & $8.4 \times 10^{-12}$ & $1.4 \times 10^{-12} \ast$   & m/W \\
$dn/dN$ & FCD coefficient & $-1.73\times 10^{-27}$ & $-1.73\times 10^{-27} $ & m$^3$ \\
$\sigma$ & FCA cross-section & $10^{-21}$ & $10^{-21}$ & m$^2$ \\
$n = n_g$ & Refractive (and group) index        & 3.476 & 3.41 $\ast$ & \\
$\eta_\mathrm{lin}$ & Absorption loss fraction & 0.4 & 0.4 & \\
$\tau_\mathrm{th}$ & Thermal relaxation time    & 65 & 15.9 $\ast$ & ns \\
$\tau_\mathrm{car}$ & Carrier relaxation time & 700 $\ast$ & 160 $\ast$ & ps \\
$\Gamma_\mathrm{th}$ & Thermal confinement factor & 0.9355 & 0.9355 & \\
$\Gamma_\mathrm{TPA}$ & TPA confinement factor  & 0.9964 & 0.9964 & \\
$\Gamma_\mathrm{FCA}$ & FCA confinement factor  & 0.9996 & 0.9996 & \\
$V_\mathrm{th}/D$ & Normalized thermal mode volume & $3.99\times 10^{-13}$ & $3.99\times 10^{-13}$ & m$^2$ \\
$V_\mathrm{TPA}/D$ & Normalized TPA mode volume & $3.24\times 10^{-13}$ & $3.24\times 10^{-13}$ & m$^2$ \\
$V_\mathrm{FCA}/D$ & Normalized FCA mode volume & $2.95\times 10^{-13}$ & $2.95\times 10^{-13}$ & m$^2$ \\
$\rho$      & Mass density of Si        & $2.33\times10^3$ & $2.33\times10^3$ & kg/m$^3$ \\
$C_p$       & Volumetric heat capacity   & 700 & 700 & J/kG/K\\
\hline\hline
\end{tabular}
\label{tab:params}
\end{table}

The next step is the incorporation of nonlinear effects via the nonlinear frequency shift $\delta \omega^\mathrm{nl}_j$ for a field labeled $j$.  By writing the nonlinear frequency shift as a complex variable, both refractive and absorptive effects can be considered.  The resulting expression contains SPM due to the probe, XPM due to the pump field $b(t)$, FCD driven by the carrier density $N(t)$, and thermo-optic effects due to to the local temperature shift $\Delta T$.  The refractive (real) part of $\delta \omega^\mathrm{nl}_j$ is given by \cite{van_vaerenbergh2012cem}:
\begin{equation}\label{eqn:domega_nl}
\mathrm{Re} \left[ \delta \omega_\mathrm{a}^\mathrm{nl} \right] = -\frac{c \omega_0 n_2 \Gamma_\mathrm{TPA}}{n_g^2 V_\mathrm{TPA}} \left( |a|^2 + 2 |b|^2 \right) - \left( \frac{\omega_0}{n_g} \frac{dn}{dN} \right) N - \left( \frac{\omega_0}{n_g} \frac{dn}{dT} \right) \Delta T.
\end{equation}
Other symbols denoting material parameters, their definitions, and values, are given in \tab{params}.  In the experiments we perform on a-Si:H, the real part plays a much stronger role in governing the observed phenomena than the imaginary part.

The  nonlinear frequency shift in \eqn{domega_nl} can then be added into the coupled wave equations for the fields $a$ and $b$ representing the probe and pump, respectively:
\begin{subequations}
\label{eqn:cwe}
\begin{align}
\der{a}{t} &= \left[ i \left( \omega_0 - \omega_a + \delta \omega^\mathrm{nl}_a \right) -\frac{\gamma_\mathrm{0}}{2} \right] a + \kappa s_\mathrm{a}(t) \label{eqn:cwe-a}\\
\der{b}{t} &= \left[ i \left( \omega_0 - \omega_b + \delta \omega^\mathrm{nl}_b \right) -\frac{\gamma_\mathrm{0}}{2} \right] b + \kappa s_\mathrm{b}(t). \label{eqn:cwe-b}
\end{align}
\end{subequations}
To fully describe the dynamics of the system, we also need dynamical equations for the evolution of the carrier density and the temperature change:
\begin{subequations}
\label{eqn:dyn-eq}
\begin{align}
\der{N}{t} &= - \frac{N}{\tau_\mathrm{car}} + \left( \frac{\Gamma_\mathrm{FCA} \beta c^2}{2 \hbar \omega_0 V_\mathrm{FCA}^2 n_g^2 } \right) |b|^4 \label{eqn:carriers}\\
\der{\Delta T}{t} &= - \frac{\Delta T}{\tau_\mathrm{th}}  + \left( \frac{\Gamma_\mathrm{th} \gamma_0}{ \rho C_p V_\mathrm{th}} \right) |b|^2 \label{eqn:temp}
\end{align}
\end{subequations}
We note that in \eqn{carriers} and \eqn{temp}, the only optical field included is $b(t)$ because it denotes the pump field which is generally several orders of magnitude stronger than the probe.

We have not discussed in detail the geometrical parameters of the microrings.  A waveguide eigenmode solver was used to simulate the modes.  We find a mode area of 0.289~$\mu$m$^2$ for the 250-nm by 450-nm waveguide cross section.  In \tab{params}, for mode volumes we list values normalized by the ring diameter such that the volumes for different diameters can be directly used.  The mode volumes $V_i$ and confinement factors $\Gamma_i$ are defined for each of the processes listed in \tab{params} \cite{barclay2005nro}.  The normalized volumes in \tab{params} are approximate as in finite-difference time-domain simulations it was found there is a slight deviation of 1\% in this scaling as the diameter $D$ goes from 5 to 10~$\mu$m.  We also note that as fitting of $n_2$ was not done rigorously, we estimate approximately $\pm25$\% error bars on our value of $n_2$ and $\beta$.

We note that in fitting the free-carrier behavior of a-Si:H there is an additional uncertainty owing to the fact that measuring the carrier effects only gives information about the product $\beta \times dn/dN$ as this is the combination that determines the FCD contribution to $\delta \omega^\mathrm{nl}_a(t)$.  We therefore chose to fix the value of $dn/dN$ to the accepted value for c-Si, which allowed us to estimate $\beta$.  In future work, we plan to do direct measurements of the FOM which should allow disambiguation of the two effects \cite{wathen2012smt}.

\section*{Acknowledgments}

This work was supported in part by the Space and Naval Warfare Systems Center Pacific and the Defense Advanced Research Projects Agency under Agreement No. N66001-12-2-4007.  We acknowledge helpful discussions with Eric Van Stryland, David Hagan, and Warren Jackson.


\begin{thebibliography}{10}
\newcommand{\enquote}[1]{``#1''}

\bibitem{beausoleil2011lsi}
R.~G. Beausoleil, \enquote{{Large-scale integrated photonics for
  high-performance interconnects},} J. Emerg. Technol. Comput. Syst.
  \textbf{7}, 6:1--6:54 (2011).

\bibitem{leuthold2010nsp}
J.~Leuthold, C.~Koos, and W.~Freude, \enquote{{Nonlinear silicon photonics},}
  Nat. Photonics \textbf{4}, 535--544 (2010).

\bibitem{harke2005lls}
A.~Harke, M.~Krause, and J.~Mueller, \enquote{{Low-loss singlemode amorphous
  silicon waveguides},} Electron. Lett. \textbf{41}, 1377 -- 1379 (2005).

\bibitem{sun2009tas}
R.~Sun, J.~Cheng, J.~Michel, and L.~Kimerling, \enquote{{Transparent amorphous
  silicon channel waveguides and high-Q resonators using a damascene process},}
  Opt. Lett. \textbf{34}, 2378--2380 (2009).

\bibitem{selvaraja2009lla}
S.~K. Selvaraja, E.~Sleeckx, M.~Schaekers, W.~Bogaerts, D.~V. Thourhout,
  P.~Dumon, and R.~Baets, \enquote{{Low-loss amorphous silicon-on-insulator
  technology for photonic integrated circuitry},} Opt. Commun. \textbf{282},
  1767--1770 (2009).

\bibitem{grillet2012asn}
C.~Grillet, L.~Carletti, C.~Monat, P.~Grosse, B.~{Ben Bakir}, S.~Menezo, J.~M.
  Fedeli, and D.~J. Moss, \enquote{{Amorphous silicon nanowires combining high
  nonlinearity, FOM and optical stability},} Opt. Express \textbf{20},
  22609--22615 (2012).

\bibitem{kuyken2011npo}
B.~Kuyken, H.~Ji, S.~Clemmen, S.~K. Selvaraja, H.~Hu, M.~Pu, M.~Galili,
  P.~Jeppesen, G.~Morthier, S.~Massar, L.~K. Oxenl{\o}we, G.~Roelkens, and
  R.~Baets, \enquote{{Nonlinear properties of and nonlinear processing in
  hydrogenated amorphous silicon waveguides.}} Opt. Express \textbf{19},
  B146--53 (2011).

\bibitem{narayanan2010bao}
K.~Narayanan, A.~W. Elshaari, and S.~F. Preble, \enquote{{Broadband all-optical
  modulation in hydrogenated-amorphous silicon waveguides},} Opt. Express
  \textbf{18}, 9809--9814 (2010).

\bibitem{matres2013hnf}
J.~Matres, G.~C. Ballesteros, P.~Gautier, J.-M. F\'ed\'eli, J.~Mart\'i, and
  C.~J. Oton, \enquote{{High nonlinear figure-of-merit amorphous silicon
  waveguides},} Opt. Express \textbf{21}, 3932--3940 (2013).

\bibitem{wang2012upc}
K.-Y. Wang and A.~C. Foster, \enquote{{Ultralow power continuous-wave frequency
  conversion in hydrogenated amorphous silicon waveguides},} Opt. Lett.
  \textbf{37}, 1331--1333 (2012).

\bibitem{kuyken2011ocp}
B.~Kuyken, S.~Clemmen, S.~K. Selvaraja, W.~Bogaerts, D.~{Van Thourhout},
  P.~Emplit, S.~Massar, G.~Roelkens, and R.~Baets, \enquote{{On-chip parametric
  amplification with 26.5 dB gain at telecommunication wavelengths using
  CMOS-compatible hydrogenated amorphous silicon waveguides},} Opt. Lett.
  \textbf{36}, 552--554 (2011).

\bibitem{wang2012upa}
K.-Y. Wang, K.~G. Petrillo, M.~A. Foster, and A.~C. Foster,
  \enquote{{Ultralow-power all-optical processing of high-speed data signals in
  deposited silicon waveguides},} Opt. Express \textbf{20}, 24600--24606
  (2012).

\bibitem{street2005has}
R.~A. Street, \emph{{Hydrogenated Amorphous Silicon}} (Cambridge University, 2005).

\bibitem{stutzmann1985lim}
M.~Stutzmann, W.~B. Jackson, and C.~C. Tsai, \enquote{{Light-induced metastable
  defects in hydrogenated amorphous silicon: A systematic study},} Phys. Rev. B
  \textbf{32}, 23--47 (1985).

\bibitem{boyd2008nlo}
R.~W. Boyd, \emph{{Nonlinear Optics}} (Academic Press, 2008), 3rd ed.

\bibitem{almeida2004aoc}
V.~R. Almeida, C.~A. Barrios, R.~R. Panepucci, and M.~Lipson,
  \enquote{{All-optical control of light on a silicon chip},} Nature
  \textbf{431}, 1081--1084 (2004).

\bibitem{turner-foster2010ufc}
A.~C. Turner-Foster, M.~A. Foster, J.~S. Levy, C.~B. Poitras, R.~Salem, A.~L.
  Gaeta, and M.~Lipson, \enquote{{Ultrashort free-carrier lifetime in low-loss
  silicon nanowaveguides},} Opt. Express \textbf{18}, 3582--3591 (2010).

\bibitem{waldow2008aos}
M.~Waldow, T.~Pl\"{o}tzing, M.~Gottheil, M.~F\"{o}rst, J.~Bolten, T.~Wahlbrink,
  and H.~Kurz, \enquote{25ps all-optical switching in oxygen implanted
  silicon-on-insulator microring resonator,} Opt. Express \textbf{16}, 7693
  (2008).

\bibitem{preston2008hsa}
K.~Preston, P.~Dong, B.~Schmidt, and M.~Lipson, \enquote{{High-speed
  all-optical modulation using polycrystalline silicon microring resonators},}
  Appl. Phys. Lett. \textbf{92}, 151104 (2008).

\bibitem{taillaert2004ceb}
D.~Taillaert, P.~Bienstman, and R.~Baets, \enquote{{Compact efficient broadband
  grating coupler for silicon-on-insulator waveguides.}} Opt. Lett.
  \textbf{29}, 2749--51 (2004).

\bibitem{little1997sri}
B.~E. Little, J.~P. Laine, and S.~T. Chu, \enquote{{Surface-roughness-induced
  contradirectional coupling in ring and disk resonators.}} Opt. Lett.
  \textbf{22}, 4--6 (1997).

\bibitem{xu2008smr}
Q.~Xu, D.~Fattal, and R.~G. Beausoleil, \enquote{{Silicon microring resonators
  with 1.5-$\mu$m radius},} Opt. Express \textbf{16}, 4309--4315 (2008).

\bibitem{niehusmann2004uqf}
J.~Niehusmann, A.~V\"{o}rckel, P.~H. Bolivar, T.~Wahlbrink, W.~Henschel, and
  H.~Kurz, \enquote{{Ultrahigh-quality-factor silicon-on-insulator microring
  resonator},} Opt. Lett. \textbf{29}, 2861--2863 (2004).

\bibitem{zhu2012eoc}
S.~Zhu, G.~Q. Lo, W.~Li, and D.~L. Kwong, \enquote{{Effect of cladding layer
  and subsequent heat treatment on hydrogenated amorphous silicon waveguides.}}
  Opt. Express \textbf{20}, 23676--83 (2012).

\bibitem{vukovic2013uoc}
N.~Vukovic, N.~Healy, F.~H. Suhailin, P.~Mehta, T.~D. Day, J.~V. Badding, and
  A.~C. Peacock, \enquote{{Ultrafast optical control using the Kerr
  nonlinearity in hydrogenated amorphous silicon microcylindrical resonators.}}
  Sci. Rep. \textbf{3}, 2885 (2013).

\bibitem{Moss2013}
D.~J. Moss, R.~Morandotti, A.~L. Gaeta, and M.~Lipson, \enquote{{New
  CMOS-compatible platforms based on silicon nitride and Hydex for nonlinear
  optics},} Nat. Photonics \textbf{7}, 597--607 (2013).

\bibitem{ref:georgas2011ald}
M.~Georgas, J.~Leu, B.~Moss, C.~Sun, and V.~Stojanovic, \enquote{{A}ddressing
  link-level design tradeoffs for integrated photonic interconnects,} in
  \enquote{{P}roceedings of the 2011 {IEEE} {C}ustom {I}ntegrated {C}ircuits
  {C}onference ({CICC}),}  (2011), pp. 1--8.

\bibitem{beausoleil2013paf}
R.~G. Beausoleil, M.~McLaren, and N.~P. Jouppi, \enquote{{Photonic
  Architectures for Data Centers},} IEEE J. Sel. Top. Quantum Electron.
  \textbf{19}, 3700109:1--9 (2013).

\bibitem{barclay2005nro}
P.~Barclay, K.~Srinivasan, and O.~Painter, \enquote{{Nonlinear response of
  silicon photonic crystal microresonators excited via an integrated waveguide
  and fiber taper.}} Opt. Express \textbf{13}, 801--20 (2005).

\bibitem{tanabe2007lpd}
T.~Tanabe, M.~Notomi, E.~Kuramochi, and H.~Taniyama, \enquote{{Large pulse
  delay and small group velocity achieved using ultrahigh-Q photonic crystal
  nanocavities.}} Opt. Express \textbf{15}, 7826--39 (2007).

\bibitem{nozaki2010sfa}
K.~Nozaki, T.~Tanabe, A.~Shinya, S.~Matsuo, T.~Sato, H.~Taniyama, and
  M.~Notomi, \enquote{{Sub-femtojoule all-optical switching using a
  photonic-crystal nanocavity},} Nat. Photonics \textbf{4}, 477--483 (2010).

\bibitem{sheik-bahae1991dob}
M.~Sheik-Bahae, D.~C. Hutchings, D.~J. Hagan, and E.~W. {Van Stryland},
  \enquote{{Dispersion of Bound Electronic Nonlinear Refraction in Solids},}
  IEEE J. Quantum Electron. \textbf{27}, 1296--1309 (1991).

\bibitem{bristow2007tpa}
A.~D. Bristow, N.~Rotenberg, and H.~M. van Driel, \enquote{{Two-photon
  absorption and Kerr coefficients of silicon for 850--2200 nm},} Appl. Phys.
  Lett. \textbf{90}, 191104 (2007).

\bibitem{vanstryland}
E.~W.~Van~Stryland and D.~J.~Hagan, Nonlinear Optics Group, CREOL, University of Central Florida, 4000 Central Florida Blvd., Orlando, FL, USA (personal communication 2013).

\bibitem{little1997mrc}
B.~Little, S.~Chu, H.~Haus, J.~Foresi, and J.-P. Laine, \enquote{{Microring
  resonator channel dropping filters},} J. Light. Technol. \textbf{15},
  998--1005 (1997).

\bibitem{van_vaerenbergh2012cem}
T.~{Van Vaerenbergh}, M.~Fiers, P.~Mechet, T.~Spuesens, R.~Kumar, G.~Morthier,
  B.~Schrauwen, J.~Dambre, and P.~Bienstman, \enquote{{Cascadable excitability
  in microrings},} Opt. Express \textbf{20}, 20292--20308 (2012).

\bibitem{wathen2012smt}
J.~J. Wathen, V.~R. Pag\'{a}n, and T.~E. Murphy, \enquote{{Simple method to
  characterize nonlinear refraction and loss in optical waveguides.}} Opt.
  Lett. \textbf{37}, 4693--5 (2012).

\end{thebibliography}
\end{document}